\documentclass[lettersize,journal]{IEEEtran}
\usepackage{amsmath,amssymb,amsfonts}
\usepackage{aligned-overset}
\usepackage{algorithmic}
\usepackage{graphicx}
\usepackage{textcomp}
\usepackage{xcolor}
\usepackage{soul}
\usepackage{textcomp}
\usepackage{stfloats}
\usepackage{url}
\usepackage{verbatim}
\usepackage{cite}
\usepackage{wrapfig}
\usepackage{subcaption}
\usepackage{hyperref}
\usepackage{gensymb}
\usepackage{tabularx}
\usepackage{longtable}
\usepackage{float}
\usepackage{soul}
\usepackage{multirow}

\usepackage{enumitem}
\usepackage{booktabs}
\usepackage{hhline}
\usepackage{boldline} 
\usepackage{ragged2e}
\usepackage{mathtools}
\usepackage{breqn}

\newcolumntype{P}[1]{>{\centering\arraybackslash}p{#1}}
\newcolumntype{M}[1]{>{\arraybackslash}m{#1}}

\graphicspath{ {./images/} }

\newcommand{\U}[1]{\mathrm{#1}}

\newcommand{\SNR}{\U{SNR}}
\newcommand{\EpsAVG}{\ensuremath{|\epsilon|_{\U{avg}}}}
\newcommand{\fs}{f_{\U{s}}}
\newcommand{\Fn}{F_{\U{n}}}
\newcommand{\Ptx}{\ensuremath{P_{\U{tx}}}}
\newcommand{\Prx}{\ensuremath{P_{\U{rx}}}}
\newcommand{\Itx}{\ensuremath{I_{\U{tx}}}}
\newcommand{\Irx}{\ensuremath{I_{\U{rx}}}}

\newcommand{\Apd}{\ensuremath{A_{\U{pd}}}}
\newcommand{\Rd}{\ensuremath{R_{\U{d}}}}
\newcommand{\gti}{\ensuremath{g_{\U{ti}}}}
\newcommand{\Vti}{\ensuremath{V_{\U{ti}}}}
\newcommand{\Vmin}{\ensuremath{V_{\U{min}}}}
\newcommand{\Vmax}{\ensuremath{V_{\U{max}}}}
\newcommand{\Vout}{\ensuremath{V_{\U{out}}}}
\newcommand{\Vnoise}{\ensuremath{V_{\U{noise}}}}
\newcommand{\rmin}{\ensuremath{r_{\U{min}}}}
\newcommand{\rmax}{\ensuremath{r_{\U{max}}}}

\newcommand{\DRmin}{\ensuremath{DR_{\U{min}}}}
\newcommand{\Bmin}{\ensuremath{B_{\U{min}}}}
\newcommand{\Bmax}{\ensuremath{B_{\U{max}}}}
\newcommand{\varT}{\ensuremath{{\sigma_{\U{T}}}^{2}}}
\newcommand{\varS}{\ensuremath{{\sigma_{\U{S}}}^{2}}}
\newcommand{\RL}{\ensuremath{R_{\U{L}}}}
\newcommand{\kB}{k_{\U{B}}}
\newcommand{\irx}{\ensuremath{i_{\U{rx}}}}
\newcommand{\Pin}{\ensuremath{P_{\U{in}}}}
\newcommand{\Pmin}{\ensuremath{P_{\U{min}}}}
\newcommand{\id}{\ensuremath{i_{\U{d}}}}

\newcommand{\commentout}[1]{}

\title{The Fundamental Limits of Light-Wave Sensing for Non-Contact Respiration Monitoring}

\author{Brenden Martin, Md Zobaer Islam, Carly Gotcher, Tyler Martinez, Sabit Ekin, \IEEEmembership{Senior Member, IEEE}, \\and John F. O'Hara, \IEEEmembership{Senior Member, IEEE}
\thanks{
This work was supported by the National Science Foundation under Grant 2008556. (\textit{Corresponding author: Brenden Martin, John F. O'Hara.})}
\thanks{Brenden Martin, Md Zobaer Islam, Carly Gotcher, Tyler Martinez, and John F. O'Hara are with the School of Electrical and Computer Engineering, Oklahoma State University, Oklahoma, USA (e-mail: zobaer.islam, brenden.martin, carly.gotcher, tyler.martinez, oharaj \{@okstate.edu\})
}
\thanks{Sabit Ekin is with the Department of Engineering Technology and Industrial Distribution, Texas A\&M University, College Station, Texas, USA (e-mail: sabitekin@tamu.edu)
}
\thanks{This work has been submitted to the IEEE for possible publication. Copyright may be transferred without notice, after which this version may no longer be accessible.}
}

\date{March 2023}

\begin{document}

\maketitle
\begin{abstract}
An experimental testbed has been constructed to assess the capabilities of Light-Wave Sensing---a promising new vitals monitoring approach.
A Light-Wave Sensing apparatus utilizes infrared radiation to contactlessly monitor the subtle respiratory motions of a subject from meters away.
A respiration-simulating robot was programmed to produce controllable, humanlike chest displacement patterns for accuracy analysis.
Estimation of respiration rate within tenths of a breath per minute has been demonstrated with the testbed, establishing the tenability of the method for use in commercial non-contact respiration monitoring equipment, and setting practical expectations on the usable range of this sensing modality.
An analytical model is then presented to guide hardware selection, and used to derive the absolute range limitations of Light-Wave Sensing.
\end{abstract}


\section{Introduction}
\IEEEPARstart{I}{f} precisely and dutifully recorded, Respiration Rate (RR) and its statistical variance can provide advanced warning of impending critical health conditions such as pneumonia, asthma attacks, pulmonary embolisms, and numerous others \cite{acute}.
Despite the value of RR as an indicator of deteriorating health, medical literature suggests that RR measurements are often made in error or forgone entirely in clinical settings \cite{neglected}.
The prevailing measurement technique of counting breaths by eye and dividing their total by elapsed time suffers from
human error and sparse monitoring \cite{van2008missed}.
Subjects who are aware of their observation experience stress and have been experimentally shown to have a statistical bias in RR of around 2 Breaths Per Minute (BPM) below that of their unaware counterparts \cite{awareness}, adding further uncertainty to RR measurements.
In light of the brevity of manual observation periods, which are on the order of a minute, it is improbable that medical personnel will be present for the most pertinent respiratory events of any given patient.
The lack of continuous, reliable respiration data is a widely recurrent cause for preventable conditions to become severe \cite{critical}.

\subsection{Existing Technology}
The plethora of issues that plague manual breath-counting have long motivated the search for automated alternatives.
Nevertheless, existing monitoring methods have failed to accrue noteworthy traction in the medical and commercial sectors to date.
These methods are reviewed and some of their shortcomings are noted.
Contact-based monitors including harnesses worn around the chest represent the most direct measurement of chest expansion.
However, they can cause discomfort to the patient and require sterilization between uses.
The invasive presence of the sensor may also impact the respiration quality of the wearer.
Respiration rate can be estimated from pulse oximetry data \cite{7748483} where photoplethysmography (PPG) is already in use, though this measurement is indirect. 
Ultrasonic sensors can be used to detect respiration from chest wall expansion, or more exotically from the Doppler shift caused by exhalation turbulence  \cite{Vortex}.
Computer vision methods have been developed, though growing privacy concerns and anti-recording sentiments foster resistance to adoption of camera-based monitoring.
Radar systems have been studied for decades in the context of vital sign monitoring \cite{1451922} \cite{2_4GHz_RADAR}, but have not seen clinical utilization.
This is in part due to system expense and the electromagnetic interference restrictions of hospital environments.
Light Detection And Ranging (LiDAR) can be used to image the entirety of a torso \cite{9438154} without the concern of interference with other medical equipment, though the hardware involved is more precise and expensive than that of radar.
Laser Doppler vibrometry has also been demonstrated as an interferometric non-contact remote monitoring (NCRM) technique \cite{Laser_Doppler_Vibrometry}, but like radar and LiDAR, demands costly sensors.
Finally, deep learning has been used to classify respiration rate with 98.69\% accuracy and patterns with 99.54\% accuracy by learning complex abstractions of Wi-Fi Channel State Information (CSI) \cite{WiFi_CSI}.
This method is unique in that it could utilize the existing Wi-Fi infrastructure, but it does not preserve the time-domain representation of the respiration signal.
Though this is not necessary for accurate classification, medical personnel may yet be able to provide additional insight if chest displacement can be viewed directly.

\subsection{Light-Wave Sensing}
The goal of this study is to quantify the fundamental limits of a new, non-contact vitals monitoring technique introduced in \cite{8935383} called Light-Wave Sensing (LWS).
LWS in general refers to detection enabled by visible and/or infrared illumination.
In this case, LWS is more narrowly defined as an approach to NCRM using a simple infrared (IR) source-detector pair to measure the chest displacements of the breathing cycle. 
In an LWS NCRM system, an incoherent, infrared beam is directed at a subject whose respiratory movements induce periodic variations in the reflected brightness.
A signal proportional to the chest motion is obtained from the torso-scattered light via photodetector.
Frequency domain analysis can then be used to isolate and track the rate information from the respiration signal.

\subsection{The Advantages of LWS}
LWS bears several characteristic features that are favorable for real-world implementation of NCRM.
The intent behind LWS NCRM is to exploit the ubiquity of infrared light to garner radar-typical results at a reduced price-point with few regulatory and privacy concerns.
The inherent safety of LWS is proven by the presence of ambient infrared in quantities which exceed the required output power for a functioning LWS system.
Privacy concerns are mitigated as the LWS data is one-dimensional, limited in scope to basic motion versus time information.
The same machine learning approaches used with other NCRM methods are equally valid to apply to LWS, though they largely constitute overkill, since the LWS signal is directly proportional to chest displacement.
When the problem of limb motion rejection is introduced, machine learning analysis becomes quite appealing, as there is spectral overlap between human respiration rates and limb movements that simple linear filtering cannot reject.
This challenge of rejecting interferers is not unique to LWS, being shared by all NCRM approaches.
However, the transparency of the respiratory information in the LWS signal suggests that shallower networks can be used with LWS data than in more abstract data such as Wi-Fi CSI.
Simple respiration monitors using LWS can be produced inexpensively.
The lower respiration monitoring costs can be driven, the more feasible deployment in low priority settings becomes.
Monitoring respiration not only in the clinic, but at home and the workplace stands to raise the likelihood of early disease detection and lessen the workloads of medical staff.
Increasing NCRM accessibility through LWS could yield greater, lifelong healthcare through the as-yet unharnessed wealth of information perpetually broadcast by the respiration cycle.

\subsection{Fundamental Limits and Outline}

Ultimately, the quality of a respiration monitoring method is assessed by the accuracy of its respiration rate estimates and ability to properly classify breathing patterns.
However, those accuracies are contingent not only on the sensing hardware, but on the nature of the breathing and the analysis technique.
As such, we seek some other quantity to describe LWS system performance---a metric irrespective of analysis and subject.
Since LWS detection is based on chest cavity motion, its estimation accuracy is a strong function of the fidelity with which chest displacement information is captured in the raw LWS signal.
To this end, we identify the range resolution of an LWS system as its most fundamental limitation.
The range resolution is the smallest chest displacement measurable by a given LWS system at a certain relative position to the subject.
By solving for the range resolution in terms of LWS hardware properties, we derive a mathematical expression that relates the system's underlying electronics to its fundamental measurement capabilities.
The utility of the expression is seen in selecting components for LWS implementation.
For example, if the range resolution of a hypothetical LWS system was found to exceed the deepest human breathing depths, then that system design would be clearly incapable of adequately capturing respiratory information.
Background clutter, limb motions, and clothing are chaotic obfuscations of the respiration signal which serve to reduce LWS accuracy.
These sources of degradation have been minimized in experimentation and omitted from analytical consideration such that the bounds provided in this paper represent absolute limits on LWS.

\label{Medical Section}

The remainder of the manuscript is organized as follows.  
The details of the experimental testbed are provided in section \ref{Testbed Section} and the RR prediction accuracy of the LWS system is analyzed in section \ref{Results Section} yielding close agreements to ground truth.
Transitioning from practical results to theoretical discourse, a modeling framework is developed in section \ref{LWS_Model_Section} as a general purpose description of LWS.
Applying some simplifications and assuming the breathing depths of normal human physiology, the model is manipulated to solve for optimal range resolution as a function of LWS system quality and subject location in section \ref{Limitations Section}.
In section \ref{Range_Resolution_Discussion}, we discuss the range resolution of our experimental system and the reduction from theoretically-limited performance caused by real environments.
To conclude, the advantages and applications of      the technology are provided in sections \ref{Application Section} and \ref{Conclusion Section}.

\section{Experimental Testbed}
\label{Testbed Section}

\subsection{Breathing Phantom}
\begin{figure}
    \includegraphics[clip, trim=0cm 0cm 0cm 0cm, width=\linewidth]{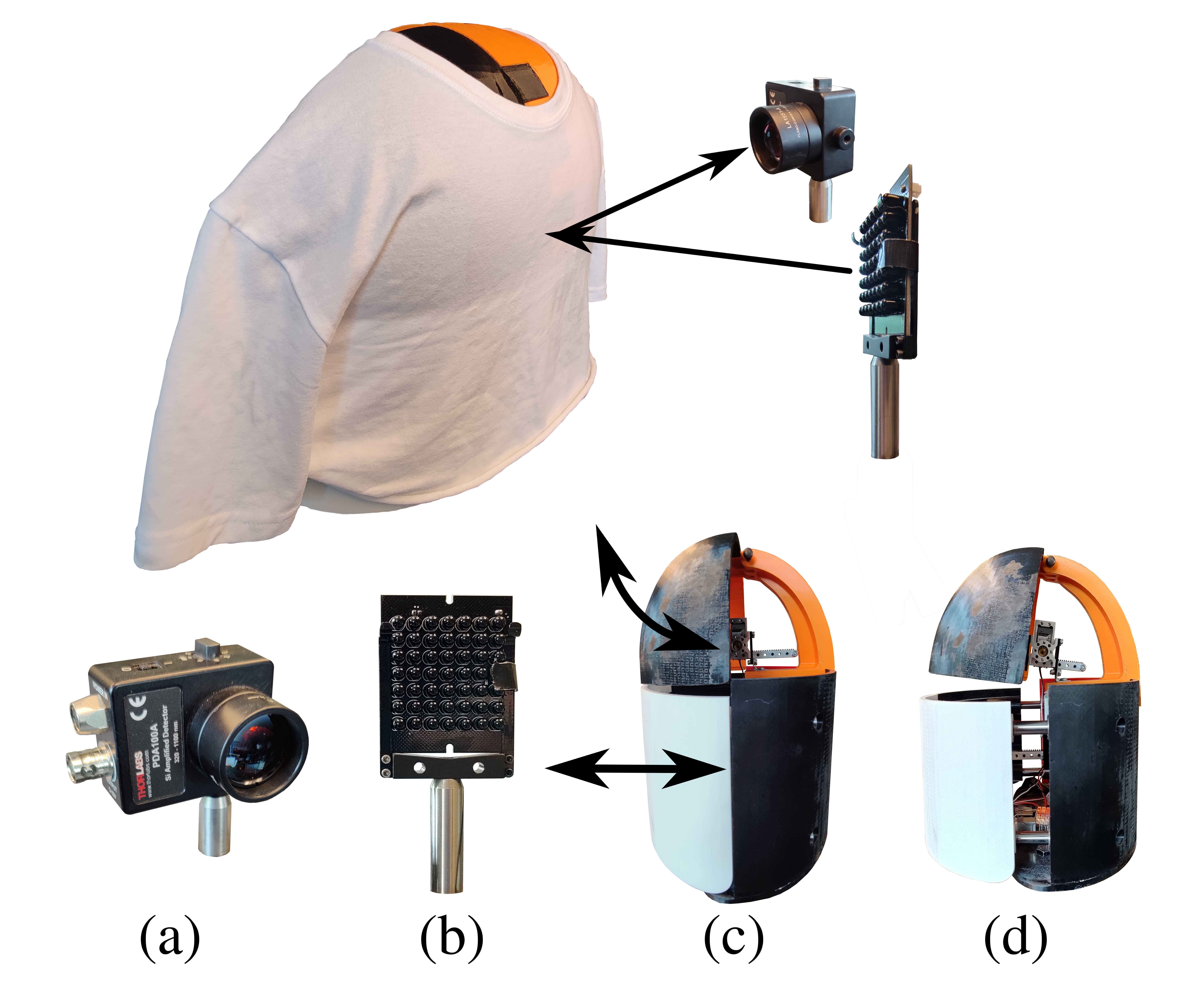}
    \caption{Function of the breathing phantom. (a) Photodetector. (b) IR array. (c) Breathing phantom actuation curves. (d) Breathing phantom at full expansion.}
    \label{Breathing_Phantom}
\end{figure}

To minimize measurement uncertainties and provide an objective ``ground truth'' upon which LWS system performance could be quantified, human subjects were excluded in this work. In their stead, an automaton surrogate known as a “breathing phantom” was constructed.
The breathing phantom, seen in Figure \ref{Breathing_Phantom}, is a two segment torso model capturing the abdominal motion of the diaphragm and the upward, angled motion of the chest and shoulders.
Had human participants been used, it would have been necessary for them to either breathe in synchronicity with a metronome or wear a contact sensor to obtain a ground truth. The breathing phantom serves as a predictable reference capable of repeatedly actuating arbitrary breathing patterns.
The maximum linear travel of the chest actuator is 30 mm, which corresponds to a typical chest expansion for deep breathing \cite{RMMI}.
The breathing phantom can simulate respiration as rapid as 35 BPM at full depth and 40 BPM at shallower depths.
These maximum rates exceed the healthy limits of resting respiration so that the breathing phantom can emulate unhealthy respiration patterns such as hyperventilation. 

\subsection{LWS System}

\begin{figure}
    \centering
    \includegraphics[width=\linewidth]{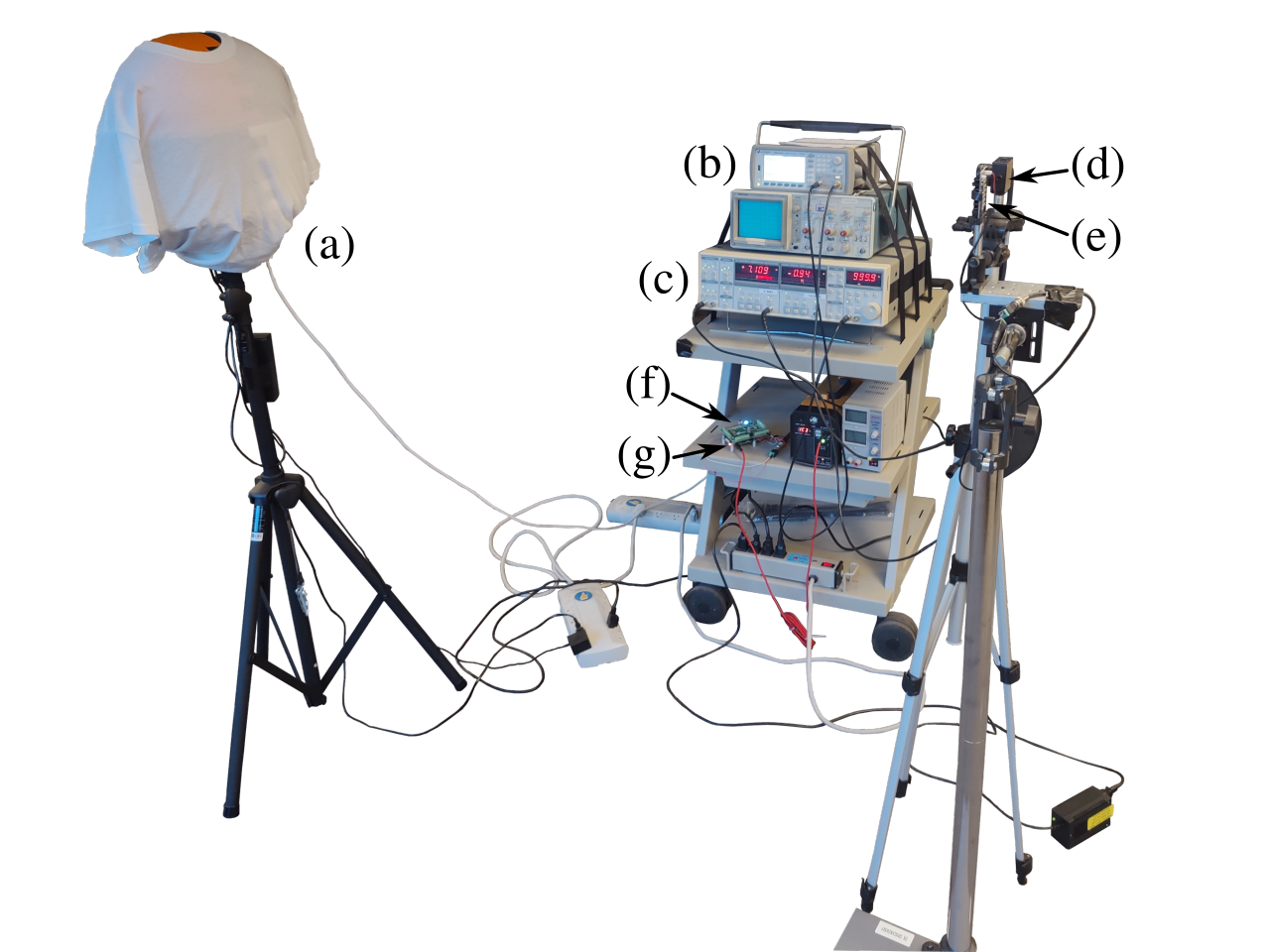}
    \caption{Experimental setup. (a) Breathing phantom. (b) Function generator. (c) Lock-in amplifier. (d) Photodetector. (e) IR LED array. (f) Analog to digital converter. (g) Microcontroller.}
    \label{Experimental Setup}
\end{figure}

\begin{figure}
    \includegraphics[clip, trim=0cm 0.5cm 0cm 0cm, width=\linewidth]{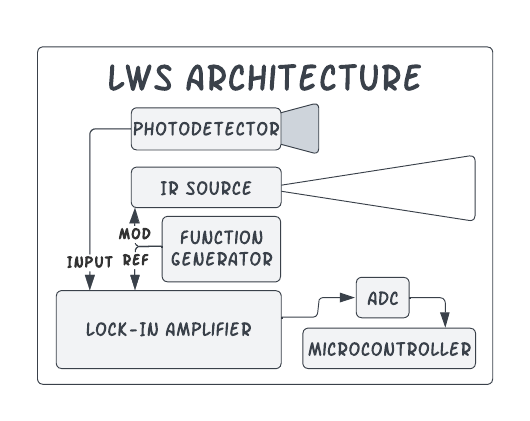}
    \caption{High-level block diagram of experimental LWS system.}
    \label{LWSBD}
\end{figure}

The LWS system itself is shown in Figure \ref{Experimental Setup}, and summarized by the architecture in Figure \ref{LWSBD}.
A ThorLabs PDA100A photodetector (\textit{p-i-n} photodiode with integrated transimpedance amplifier) was used to collect the light scattered off the subject (i.e. the clothed phantom).  The source was an array of 48 IR LEDs.  The array was driven at 1 kHz by a Keysight 33500B function generator to enable lock-in detection with a Stanford Research Systems SR830 lock-in amplifier.
A Pi-Plate analog-to-digital converter (ADC), in conjunction with a Raspberry Pi 4, serves to sample and store output values from the lock-in amplifier.
Lock-in amplification isolates the modulated LWS beam from the ambient infrared background.
The lock-in amplifier produces a quasi-DC output which is proportional to the 1 kHz modulated light only.
The harmonics of limb motion at 1 kHz are negligible and there are no notable 1 kHz optical interferers in general home or clinical settings.
In this way, the LWS system is highly selective of its own transmit beam even though the photodetector has a wideband response.
The time constant for lock-in amplification $\tau = 100$ ms was selected to minimize noise as much as possible while maintaining the fine temporal details of the breathing motions.

\commentout{
\hl{The longer the time constant $\tau$ used in low-pass filtering, the more selective the lock-in detection and the more resilient the system is to noise.
$\tau$ must be short enough to allow the lock-in amplifier to respond to the highest respiration rates of around 35 BPM, which corresponds to about 0.6 Hz.
The cutoff frequency for a low-pass filter is often defined in terms of its time constant by $f_c = \frac{1}{2 \pi \tau}$.
Setting $f_c = 0.6$ yields $\tau = \frac{1}{2 \pi f_c} = 265$~ms.
By this definition of cutoff, the power is attenuated to half at $f_c$.
However, to ensure approximately no attenuation up to this frequency, the time constant was restricted to $\tau \leq 100$~ms, providing a reasonably flat frequency response. (Again maybe too much info depending on the journal.  This could be condensed to one sentence. The lock-in technique is very well known.)}
}

\section{Experimental Results}
\label{Results Section}


\begin{figure*}
\includegraphics[clip, trim= 2.2cm 0cm 2.5cm 0cm, width=\textwidth]{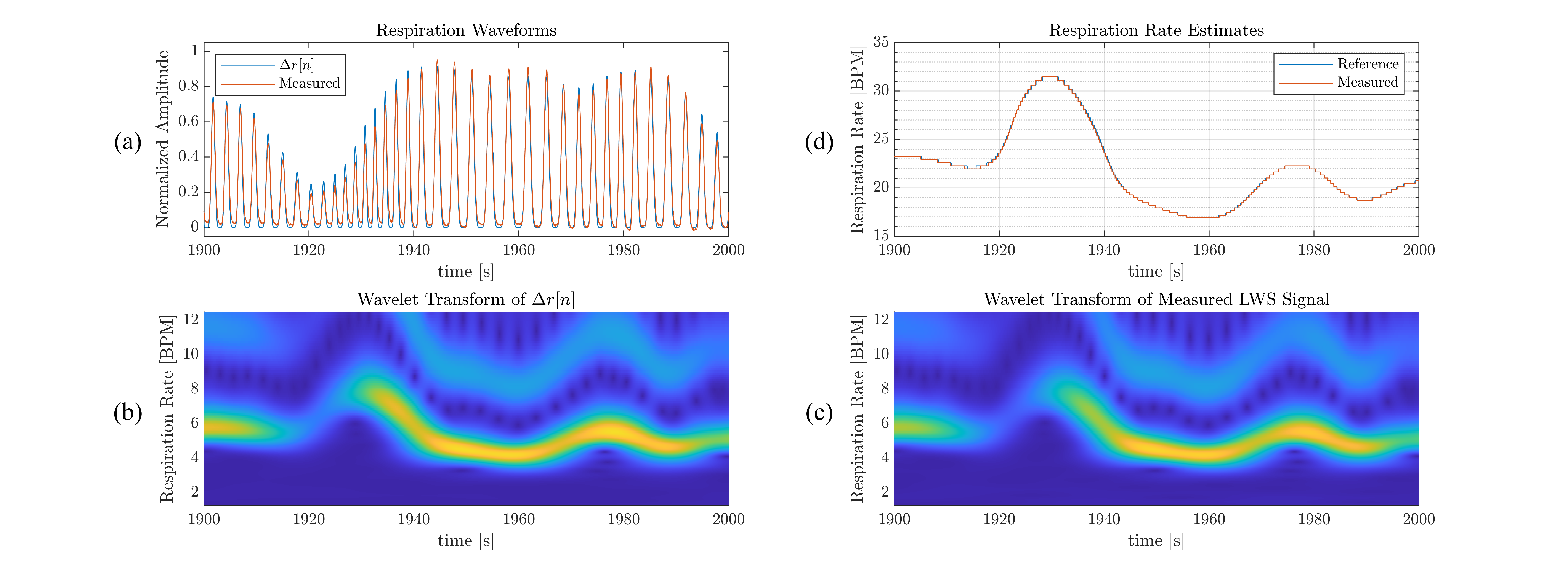}
\caption{Respiration rate analysis. The distance between the LWS system and breathing phantom $r_0 = 1$ meter. (a) The ground truth chest displacement $\Delta r[n]$ commanded to the breathing phantom plotted alongside the measured LWS signal. The bottom insets are the wavelet transforms of (b) $\Delta r[n]$ and (c) the measured LWS signal. (d) Instantaneous respiration rate estimates computed from $\Delta r[n]$ and the measured LWS signal.} 
\label{stochastic_fig}
\end{figure*}

As proof-of-concept, we have applied wavelet-based rate detection to data collected from our LWS system and breathing phantom.
A pure sinusoidal signal can be extracted from an excessively high noise floor given a long enough Fourier transform window.
Genuine human respiration waveforms, in contrast with the sinusoid, possess stray harmonic content and unpredictable variations in rate which pose additional challenges to analysis.
To evaluate the LWS system performance in the context of its intended use case, this accuracy analysis has been performed using stochastic breathing waveforms modeled after healthy and unhealthy human respiration statistics.
First, let us discuss the wavelet respiration rate detection method.

\subsection{Wavelet Transform}
The wavelet transform (WT) uses short pulse packets called wavelets as basis functions to decompose a signal into its spectral content as a function of time.
It provides enhanced time and frequency localization of short pulses compared to the fast Fourier transform (FFT) or short-time Fourier transform (STFT).
Hence the wavelet transform is well-tailored for vitals monitoring for which non-stationary characteristics are of special importance.
A wavelet filter bank spanning the frequency range of human respiration is applied to the breathing waveform.
The frequency corresponding to the most excited wavelet filter is identified as the instantaneous respiration rate.
The binning of the WT results in a discrete spectrum of potential RR estimates.
The spacing between WT frequency bins is inherently nonlinear.
The density of bins should be carefully selected so the worst case resolution is still acceptable.
In the following accuracy analyses 48 voices per octave were chosen providing a respiration rate resolution of 0.7 BPM at 10 BPM and 0.15 BPM at 40 BPM.
It is possible to attain less discretized RR estimates by comparing several filters in the frequency neighborhood around the maximum, and weighted-averaging each of their frequencies by their relative excitation amplitudes.
However, this does not significantly impact accuracy results lending to the preference of the straightforward maximum for the following analysis.

\subsection{Stochastic Respiration Waveforms}
Now we turn our attention to the design of humanlike respiration patterns.
The respiration patterns were generated as a discrete time series at the sampling rate $\fs=100$~Hz, in accordance with the breathing phantom refresh rate.
The normalized chest displacement pattern $\Delta r[n]$ sent to the breathing phantom is computed as:
\begin{equation}
  \Delta r[n] = A[n] \sin^6(\phi[n]), \quad n = 0,1,2,...
  \label{Displacement_Formula}
\end{equation}
where \(A[n]\) is a random variable performing amplitude modulation and \(\phi[n]\) is the instantaneous phase.
The use of pure sines to represent respiration is common in NCRM literature, however it fails to capture the asymmetry of inspiration and expiration periods.
The use of evenly-exponentiated sines to incorporate harmonics beyond the fundamental breathing frequency has been argued for radar applications \cite{Breathing_Model}.
Thus sine is raised to the power of 6 in this model to better match the profile of human breathing.
The amplitude envelope $A[n]$ is created by applying the arctangent to correlated Gaussian noise $a[n]$ to definitely bound the envelope:
\begin{equation}
   A[n] = d_{\U{min}} + \left [ \frac{\arctan(a[n])}{\pi} + \frac{1}{2} \right ] (1-d_{\U{min}}).
\end{equation}

As a result, $d_{\U{min}} < A[n] < 1$.
If the fractional minimum breathing depth is selected in the range \(0 < d_{\mathrm{min}} < 1\), there will be breathing present everywhere in the signal.
If \(d_{\U{min}} = 0\), apnea conditions can be modeled.
To match the physical situation of breathing, the amplitude envelope should be approximately constant for the duration of each breath.
Accordingly, random variable $a[n]$ has a correlation time $\tau_c$ whose duration is longer than the slowest full breathing period.

The instantaneous phase in equation \ref{Displacement_Formula} can be computed from an instantaneous frequency sequence \(f[n]\).
As with the amplitude envelope, it is not physically meaningful for the respiration rate to vary during a single breath.
Therefore, the correlation time \(\tau_c\) used for \(a[n]\) is also used in generating the frequency random variable \(f[n]\).
Over the course of each sampling period \(\Delta t = \frac{1}{\fs}\), the phase increments by \( \Delta\phi[n] = 2\pi f[n] \Delta t\).
The instantaneous phase is the accumulation of the incremental phase up to the current sample:
\begin{equation}
    \phi[n] 
    = \frac{1}{2} \sum_{k=0}^{n} \Delta \phi[k]
    = \pi \sum_{k=0}^{n} f[k] \Delta t .
\end{equation}

The factor of $\frac{1}{2}$ is necessary because there are two pulses per each $2\pi$ period of $\sin^6(\phi[n])$.

\subsection{Accuracy Analysis}
\label{Accuracy Subsection}
Measurements of stochastic respiration waveforms were conducted in stretches of one hour.
The breathing phantom was placed at ranges $r_0$ from 0.5 to 2.5~m in 0.5~m increments.
At each distance, two classes of breathing were measured.
The first respiration pattern was engineered with a rate distribution typical of a healthy adult.
The second pattern was given elevated respiration rates of increased standard deviation---traits which are often linked to unhealthy breathing.
The mean and variance of $f[n]$ for the two patterns were selected in accordance with known realistic values \cite{acute}. 
Amplitude modulation with $d_{\U{min}} = 0.3$ was used so that periods of both shallow and deep breathing are included in each pattern.
Over the course of an hour, more than a thousand respiratory cycles occur, covering a diverse ensemble of respiration rates and amplitudes.
To enforce self-consistency and repeatability on the data, the same two respiration patterns were measured at each distance.
This methodology was intended to ensure the relative accuracies per distance are a meaningful comparison.

The wavelet rate detection technique outlined earlier in this section was applied to both the reference waveform and the measured signal from the LWS system.
Figure \ref{stochastic_fig}(a) directly illustrates the strong similarity of the reference and normalized LWS signal in the time domain at distance \(r_0 = 1\) meter.
The WTs of the reference and measured signal are provided in Figure \ref{stochastic_fig}(b) and \ref{stochastic_fig}(c) respectively.
Extracting the fundamental frequencies from the two wavelet spectra yields the instantaneous respiration rates plotted in Figure \ref{stochastic_fig}(d).
The excellent ability of the LWS system to estimate respiration rates even during rapid breathing transitions is visually apparent in Figure \ref{stochastic_fig}(d), in which the reference and measured rates almost entirely overlap.
The absolute difference in the two RR estimate signals is averaged over time to yield:
\begin{equation}
\EpsAVG = \left < | RR_{\U{measured}} - RR_{\U{hypothetical}} | \right>.
\end{equation}

The following table conveys the \(\EpsAVG\) of our experimental testbed up to 2.5 meters.

\begin{center}
\label{static}
\begin{tabular}{||c | c | c||} 

 \hline
 \(r_0\) & Healthy \(\EpsAVG\) & Unhealthy \(\EpsAVG\) \\ [0.5ex] 
 \hline\hline
 0.5 m &  0.019 BPM & 0.043 BPM\\ 
 \hline
 1 m &  0.022 BPM & 0.036 BPM\\
 \hline
 1.5 m &  0.022 BPM & 0.068 BPM\\
 \hline
 2 m &  0.052 BPM & 0.086 BPM\\
 \hline
 2.5 m &  0.200 BPM & 0.605 BPM\\
\hline
 
\end{tabular}
\end{center}

The LWS system preserves the relevant spectral content of the reference waveform faithfully up to 2.5 meters away, with a rapid decline in performance for farther distances.
It is true that swaying motions and other random interferers will yield reduced accuracies when LWS is used to measure human subjects.
However, these same problems are shared by all NCRM technologies.
More importantly, the accuracy results of this section demonstrate that LWS is capable of capturing chest displacement information with minimal distortion.

\section{The LWS Model}
\label{LWS_Model_Section}
\begin{figure}
    \centering
    \includegraphics[clip, trim = 6cm 13cm 6cm 12cm, width=\linewidth]{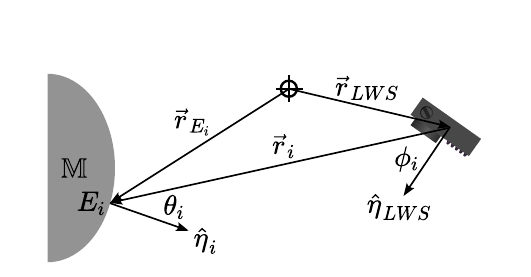}
    \caption{Collocated, codirectional LWS model geometry.}
    \label{Model Geometry}
\end{figure}
The following model describes the LWS return signal from a fabric scattering surface.
The model can be used to predict the LWS signal strength for various body sizes, breathing depths, subject positions, and hardware selections.
The model is separated into three components encapsulating the traits of the source, scatterer, and detector.
Each will be discussed individually, but first it is useful to define the objects and geometry of the model (Figure \ref{Model Geometry}).
We represent the clothed torso with mesh \(\mathbb{M}\) containing fabric scattering elements \(\mathbb{M} = \{E_0,E_1,...,E_N\}\).
\(\mathbb{M}\) is illuminated by the source, and the light scattered from each element contributes to the optical power measured at the detector.
Each element is defined by a position vector, normal unit vector, area, and a material absorption parameter. 
Respectively \(E_i=\{\Vec{r}_{E_i},\hat{\eta}_i,A_i,\alpha_i\}\).
\(\mathbb{M}\) is selected such that no element shadows another, representing only the portion of the surface with a line of sight to the LWS system.
Consider the environment as clutterless, so that the entirety of the received signal is due to fabric elements \(E_i\) of \(\mathbb{M}\).
For this model, we assume that the distance between the source and detector is small enough that they may be handled as collocated and codirectional.
In this overlapping treatment, their shared position vector is $\Vec{r}_{\U{LWS}}$ with unit normal vector $\hat{\eta}_{\U{LWS}}$.
The distance between the LWS system and scatterer $E_i$ is $r_i = |\Vec{r}_i| = |\Vec{r}_{E_i}-\Vec{r}_{\U{LWS}}|$.
The angle between the surface of $E_i$ and its ray of illumination is $\theta_i = \arccos{(-\frac{\Vec{r}_i}{r_i} \cdot \hat{\eta}_{i})}$.
The angle to the scattering element with respect to the LWS system is $\phi_i = \arccos{(\frac{\Vec{r}_i}{r_i} \cdot \hat{\eta}_{\U{LWS}})}$.
The contribution of each element $E_i$ is calculated individually with the total return signal obtained as the sum of the elemental contributions. 

\subsection{Source}
\begin{figure}
    \includegraphics[clip, trim = 1.25cm 0.5cm 0.5cm 2cm, width=\linewidth]{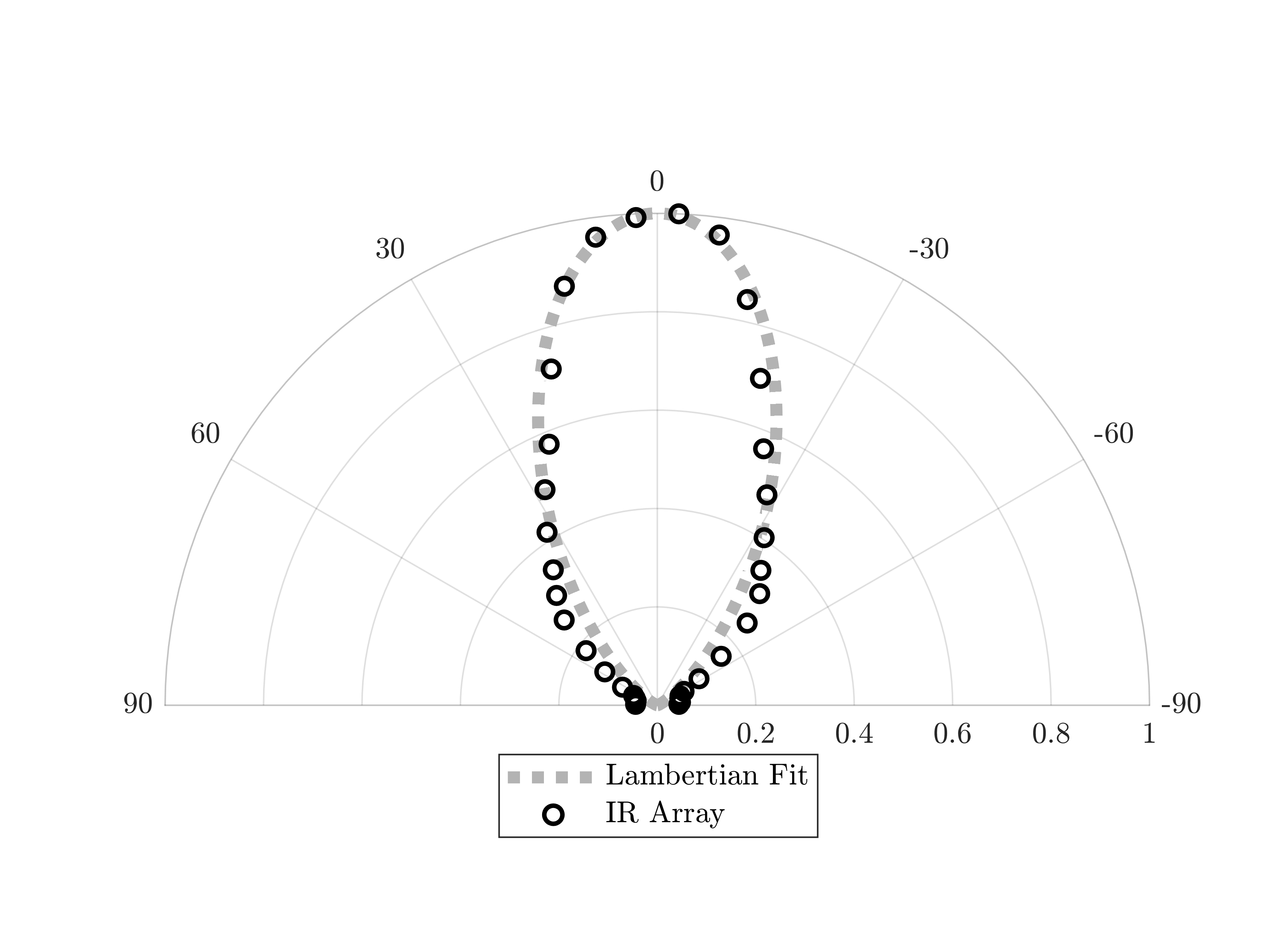}
    \caption{Source intensity profile. The measured intensities of the IR array used in this study are superimposed on a Lambertian fit with order $n = 5.78$.}
    \label{Lambertians}
\end{figure}

Lambertian intensity profiles are mathematically well-governed and are established in optics literature as a reasonable fit for modeling light-emitting diodes (LEDs) \cite{7239528}\cite{1455777}.
The Lambertian model assumes the total optical power of the source $P_{\U{tx}}$ is radiated from a point.
At distance \(r_i\)
from the source point with an off-normal angle \(\phi_i\), the intensity from a Lambertian source is given by
\begin{equation}
\Itx (r_i,\phi_i)=\frac{(n+1)\Ptx}{2\pi r_i^2}\cos^n(\phi_i). 
\end{equation}

The generalized Lambertian profile consists of a single lobe, whose shape is controlled entirely by the Lambertian order parameter, $n$.
The Lambertian order can be determined from the half power semiangle  $\phi_{\frac{1}{2}}$ according to
\begin{equation}
\label{order}
n=\frac{-\ln{2}}{\ln{\cos\phi_{\frac{1}{2}}}}.
\end{equation}

The scattered points in figure \ref{Lambertians} are normalized intensity samples taken as a function of angle from our LWS system's IR array.
Even though this source is a grid of LEDs rather than a single LED, the LEDs are closely distributed.
In the relevant ranges for LWS, the difference in LED position is negligible and the array acts as a point source.
The Lambertian profile with order $n = 5.78$ is well-matched to observed behavior, and is plotted in figure \ref{Lambertians} for comparison.

\subsection{Scatterer}

\begin{figure}
    \includegraphics[clip, trim = 1.25cm 0.5cm 0.5cm 2cm, width=\linewidth]{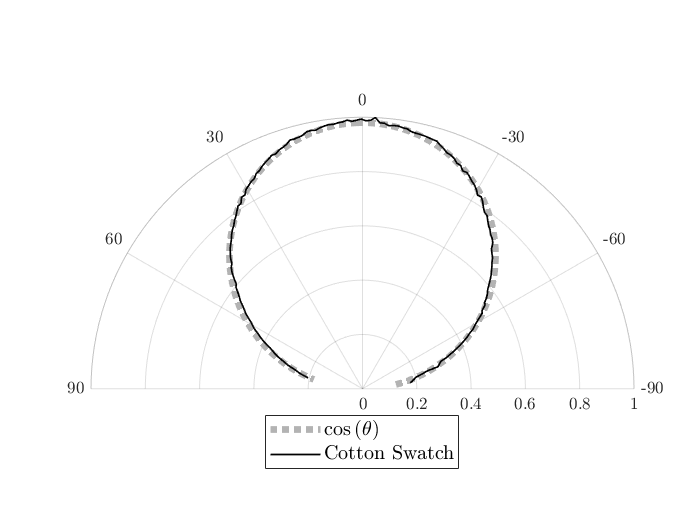}
    \caption{Normalized cotton intensity scattering profile. The measured light scattering profile from a swatch of cotton and the ideal Lambertian profile, $n = 1$ (i.e. $\cos{(\theta)}$).}
    \label{Cotton}
\end{figure}


If $\mathbb{M}$ is a sufficiently fine mesh, the optical intensity over the surface of each individual element can be considered constant.
In that case, \(E_i\) intercepts power \(P_{E_i}=I(r_i,\phi_i)\cos(\theta_i)A_i\).
The fabric element is modeled as a purely diffuse scatterer, with a normalized scattering profile of $\cos(\theta_i)$.
To test the validity of this description, a laser was used to confine the entirety of the signal power to a small patch of a cotton t-shirt wrapped around a flat panel.
This emulates the ray optics approximation of the model by grouping the entirety of the power at a point.
The scattered intensity levels provide the angular distribution of re-radiated light per \(P_{E_i}\).
The adherence of the theoretical model to the rotational scattering profile obtained experimentally is exemplified in Figure \ref{Cotton}, justifying the selection of the Lambertian scattering profile to represent the scattering from the cloth elements of \(\mathbb{M}\).
The scattered intensity at the detector is then:

\begin{equation}
\label{I_rx}
I_{\U{rx}} = \frac{\alpha_i P_{E_i}}{2\pi r_i^2} \cos(\theta_i) =  \frac{\alpha_i I_{\U{tx}}(r_i,\phi_i) A_i}{2\pi r_i^2} \cos^2(\theta_i).
\end{equation}

Representing hemispherical re-radiation, the additional freespace pathloss term \(\frac{1}{2\pi r_i^2}\) is incurred.
The material absorption parameter \(\alpha_i\) accounts for the power dissipated in the fabric.
Our trials have been conducted using 95\% cotton 5\% polyester blend t-shirts for their prevalence.
A white swath illuminated by a source with wavelength \(\lambda = 800\)~nm illumination was determined to have \(\alpha_i = 0.103\) from its measured scattering profile.
A handful of fabric colors were investigated having \(0.07 \lesssim \alpha_i \lesssim 0.1\), so \(\alpha_i \approx 0.1\) is taken to be a rule-of-thumb in the near-IR for performance estimation.

\subsection{Detector}
In a field of photon flux, a semiconductor slab impinged at angle $\phi_i$ has an effective area of $A_{\U{eff}} = A_{\U{pd}} \cos(\phi_i)$, where $A_{\U{pd}}$ is the aperture area of the photodiode.
To accommodate for the field of view (FOV) of a lens, $A_{\U{eff}}$ can be modified to $A_{\U{pd}} \cos^m(\phi_i)$, where $m$ is fractional for expanded FOVs and $m>1$ for tighter FOVs.
The order parameter $m$ can be calculated from the half power angle of the detector profile just as $n$ was determined from the source profile using equation \ref{order}.

Thus, generally, the power intercepted by the detector is $\Prx = A_{\U{eff}} \Irx = A_{\U{pd}}\cos^m(\phi_i)I_{\U{rx}}$, which is then transformed into current and voltage signals by the measurement instrumentation. 
The responsivity of the photodiode \(R_{\U{d}}\) relates the incident optical power to the signal photocurrent by \(\irx = \Rd \Prx\).
The responsivity is wavelength dependent, so in rigorous treatment of a wideband source, one must integrate the product of the optical power spectrum and responsivity spectrum.
Here, however, \(\Rd\) is assumed to be roughly constant over the LED bandwidth.
The photocurrent is fed into a transimpedance amplifier, producing a voltage signal \(\Vti = \gti \irx\), where \(\gti\) is the transimpedance gain.
The lock-in amplifier receives this voltage and amplifies it to a full scale output voltage, \(V_{\U{full}}\), determined by the sensitivity setting $S$.
The final voltage recorded by the microcontroller, \(\Vout\), is then given by:
\begin{equation}
\label{V_out_proportion}
\Vout =\frac{V_{\U{full}}}{S} \gti \Rd \Prx.
\end{equation}

\subsection{The LWS Equation}
Collecting the individual expressions for source, scatterer, and detector, we propose the following model for the LWS return signal, the intercepted power due to the scattering from mesh \(\mathbb{M}\):

\begin{equation}
\Prx = \sum_{i=0}^{N} {\frac{\alpha_i(n+1) \Ptx \Apd A_i}{4\pi^2r_i^4}\cos^{m+n}(\phi_i)\cos^2(\theta_i)}.
\label{LWS}
\end{equation}

Due to clutter, occlusions, and multipath propagation within a particular environment, the freespace treatment, having \(\Prx \propto r_i^{-4}\), may not adequately represent the LWS optical channel.
Borrowing from wireless communications theory \cite{Goldsmith}, we can replace the idealistic freespace pathloss exponent of \(4\) with an empirical pathloss exponent \(\gamma\) and generalize the LWS equation into the following form:

\begin{equation}
\Prx = \sum_{i=0}^{N} {\frac{\alpha_i(n+1) \Ptx \Apd A_i}{4\pi^2} K \left[ \frac{\rho}{r_i} \right]^\gamma \cos^{m+n}(\phi_i)\cos^2(\theta_i)}.
\label{LWS_EQN_modified}
\end{equation}

The additional fit parameters $K$ and $\rho$ are used with $\gamma$ to match the calculated $\Prx$ to calibration measurements of $\Prx$ within the channel to be modeled.
The fit model can be used to account for the nonidealities of a subject's surroundings and provide more realistic descriptions of LWS behavior in characterized channels.

\subsection{Rotational vs Translative Motion}
\label{Rotational}
\begin{figure}
    \includegraphics[clip, trim=3cm 8.5cm 3cm 8.6cm, width = \linewidth]{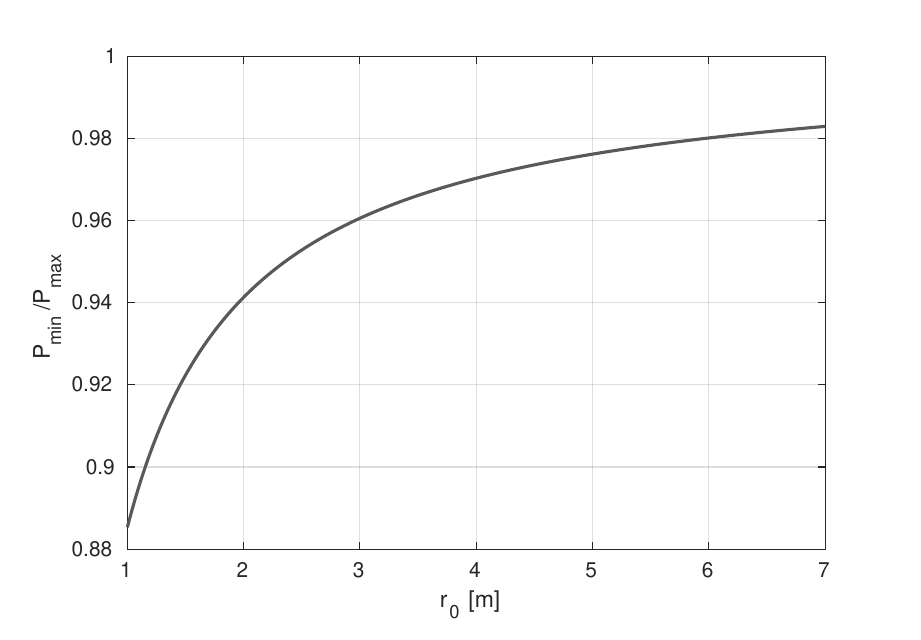}
    \caption{Relative magnitude of theoretical minimum to maximum received optical power for a chest displacement of \(\Delta r = 30 \)~mm over varying measurement distances $r_0$.}
    \label{RelAmp}
\end{figure}

Having established a general model for LWS systems, we will now apply the model to explain the falloff in LWS performance experimentally observed when measurement range $r_0$ exceeded 2.5 meters.
For illustration, let us consider the single element mesh \(\mathbb{M} = \{E_0\}\).
The torso \(\mathbb{M}\) is set at a distance \(r_0\) from the LWS system.
Suppose that the LWS system and \(E_0\) are aligned so that they face each other directly: \(\hat{\eta}_{LWS} = -\hat{\eta}_{E_0}\).
To represent chest expansion, fabric scattering element \(E_0\) is translated toward the LWS system by \(\Delta r\).
As \(E_0\) approaches the LWS system, the received power transitions from \(P_{\U{min}}\) to \(P_{\U{max}}\).
The rudimentary choice of a single element \(\mathbb{M}\) obviates the spatial distribution of the torso, which here acts as a point scatterer, having \(\phi_0=0\) and \(\theta_0=0\) in both positions.
Though this seems an over-simplification, the notion that \(\phi_i\) and \(\theta_i\) undergo relatively small variations from their initial values within the relevant ranges holds true for more complicated torso models, so long as \(\Delta r \ll r_0\).
These simplifying assumptions enable us to make an important general observation about the comparative effects on the LWS signal of rotating and translating \(\mathbb{M}\).
From equation \ref{LWS}, 
\(\Prx \propto 1/r_i^4\), so we may express the ratio of minimum to maximum received power as
\begin{equation}
\label{Relative Amplitude}
\frac{P_{\U{min}}}{P_{\U{max}}}=\left[\frac{r_0-\Delta r}{r_0}\right]^4,
\end{equation}
which is plotted for \(\Delta r = 30 \)~mm in Figure \ref{RelAmp} for \(1 \leq r_0 \leq 7\) m.
As the average distance to \(\mathbb{M}\) is increased, \(P_{\U{min}}\) approaches \(P_{\U{max}}\).
Thus the proportional variation the signal due to translation is reduced with increasing range.

The relative change imparted by rotation, contrarily, is independent of range, which can be seen from equation \ref{I_rx}.
As such, we observe in practice the bifurcation into distinct far and near LWS behaviors.
Near the LWS, the pathloss differences are substantial; far from the LWS, the pathloss varies so little that rotational motion dominates the signal.
That is, the trade-off for the simplicity of the LWS system is that it precludes itself from functioning predictably at large distances, where the perturbance of a wrinkle is as relevant to the intensity as the net expansion of the chest wall. 
While rotational disturbances at large distances obscure the fine details of the respiration waveform, they are nevertheless modulated by the respiration cycle. 
Therefore the rotational disturbances possess frequency components at the breathing rate and do not completely exclude the possibility of rate detection. 

\section{Fundamental Limitations of LWS}
\label{Limitations Section}
The most fundamental electronics limitation on the performance of an LWS system is the dynamic range of the detector.
The dynamic range is the ratio of the maximum measurable input signal value to the smallest detectable signal change.
The dynamic range of a voltage signal \(V\) is expressed in decibels (dB) as:

\begin{equation}
\label{Dynamic_Range_Formula}
DR = 20\log_{10} \left ( {\frac{\Vmax}{\Delta \Vmin}} \right ).
\end{equation}


At a range \(r_0\), the minimum measurable chest displacement \(\Delta \rmin\) is set by the dynamic range of the LWS system.
If the LWS signal is processed digitally, \(DR\) is directly related to \(B\), the number of Analog to Digital Converter (ADC) bits used to record the signal.
However, one cannot achieve arbitrary precision by adding bits indefinitely.
The useful dynamic range is hard-limited to the dynamic range of the signal itself.
Ultimately, then, the range resolution is determined by either the number of ADC bits or the detector noise floor, whichever limiting factor is more restrictive.


\begin{figure}
    \includegraphics[clip, trim=3cm 8.5cm 2cm 8.5cm, width = \linewidth]{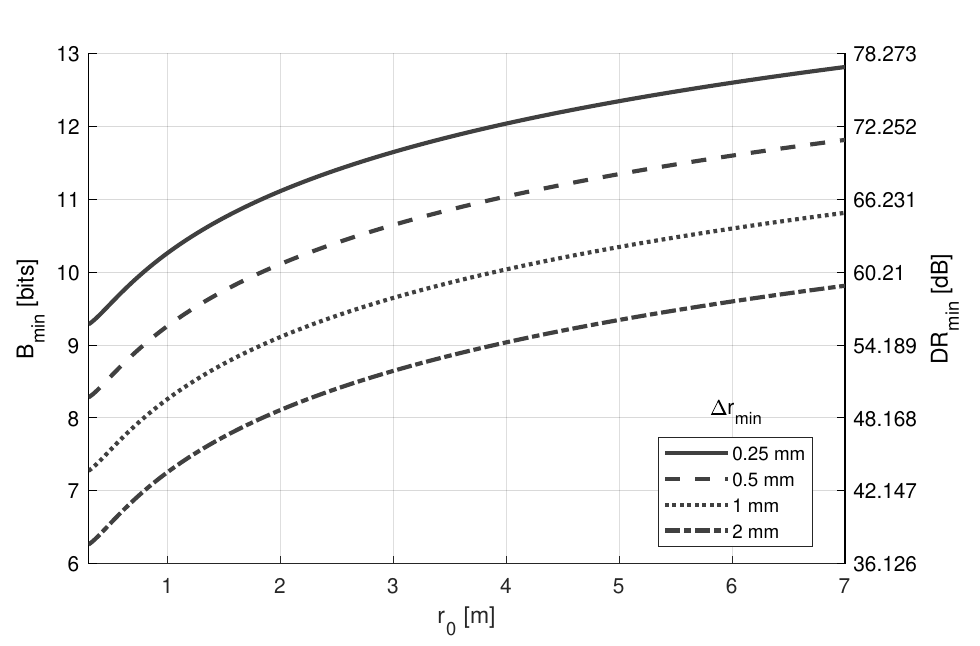}
    \caption{Minimum dynamic range for various desired $r_0$ and $\Delta \rmin$. The left vertical scale conveys the dynamic range in terms of bits. The equivalent decibel representation is labeled on the right vertical scale.}
    \label{Dynamic Range}
\end{figure}

\subsection{ADC Bits Limitation}
We first assume noiseless conditions so that the range resolution is set purely by the number of ADC bits used in capturing the signal.
The single element torso model of subsection \ref{Rotational} is overidealized, however we will employ it here to set an absolute bound on the capabilities of LWS.
The resulting limits should not be presumed realizable with genuine human subjects, though they do grant realistic expectations of the hardware needed for LWS system design.
Again, assuming \(\theta_0 = 0\) and \(\phi_0 = 0\), the input voltage to the lock-in amplifier ADC is :
\begin{equation}
    \Vti(r) = \gti \Rd \frac{\alpha_i(n+1) \Ptx \Apd A_i}{4\pi^2r^4}.
\end{equation}

The proximity of the subject to the LWS system at some time during the breathing motion is \(r (t) = r_0 - \Delta r (t)\).
Over the course of the breathing motion, \(0 \leq \Delta r (t) \leq \Delta r_{\U{max}}\).
It is sufficient for the following derivation simply that \(\Vti(r) \propto r^{-4}\). 
As a consequence of \(\Delta r \ll r_0\), \(\Vti(r)\) is well-suited to linearization in the practically pertinent regions.
If the lock-in amplifier sensitivity $S$ is optimally set---an appropriate assumption for a theoretical limit---then the input voltage signal resulting from the maximum displacement \(\Delta \rmax\) should occupy the full scale of the input range, thus:
\begin{equation}
    \Vmax = \Vti(r_0 - \Delta \rmax).
\end{equation}

Additionally, translation of the target by \(\Delta \rmin\) must vary \(\Vti\) by one LSB (least significant bit) of the input ADC so that:
\begin{equation}
    \Delta \Vmin = \Vti(r_0-\Delta \rmin) - \Vti(r_0)\
\end{equation}

\noindent and

\begin{equation}
    \label{Bits_Range_Eqn}
    \begin{split}
        \frac{\Vmax}{\Delta \Vmin} &= \frac{ \Vti(r_0-\Delta \rmax) }{ \Vti(r_0-\Delta \rmin) - \Vti(r_0) }
        \\
        &= \frac{(r_0-\Delta \rmax)^{-4}}{(r_0-\Delta \rmin)^{-4}-r_0^{-4}}.
    \end{split}
\end{equation}

The minimum number of ADC bits needed to capture a displacement of \(\Delta \rmin\) is:

\begin{equation}
    \label{B_min}
    \Bmin =  \left \lceil { \log_2{ \frac{\Vmax}{\Delta \Vmin} }} \right \rceil .   
\end{equation}

The brackets in equation \ref{B_min} represent rounding up, since the number of bits in the system must be an integer.
Noting that the formulae for bits and dynamic range (in dB) differ only in the base of the logarithm used, we can also express the minimum bits as an equivalent minimum dynamic range:

\begin{equation}
    \label{DR_min_B}
    \DRmin = \frac{20}{\log_{2}(10)} \Bmin = 6.021 \Bmin.
\end{equation}





This approach has merely quantified the smallest number of ADC bits required to perceive a minimum chest displacement under the condition that the LWS signal is received for all admitted chest positions without saturation.
According to one available dataset \cite{RMMI}, including 100 healthy males and females between ages 20 and 69, \(\Delta \rmax\) = 50.29~mm.
The shallowest abdominal breathing measured in the same work was 2.33~mm, with an average shallow breathing between men and women of 6.87~mm.
Plotted in Figure \ref{Dynamic Range} are the bit requirements for several \(\Delta \rmin\) in the range of interest. 
As expected, the required number of bits increases with the operating range $r_0$ and/or when finer chest displacement measurements $\Delta \rmin$ are selected.

\begin{figure}
    \centering
    \includegraphics[clip, trim=4cm 8.8cm 3cm 9cm, width = \linewidth]{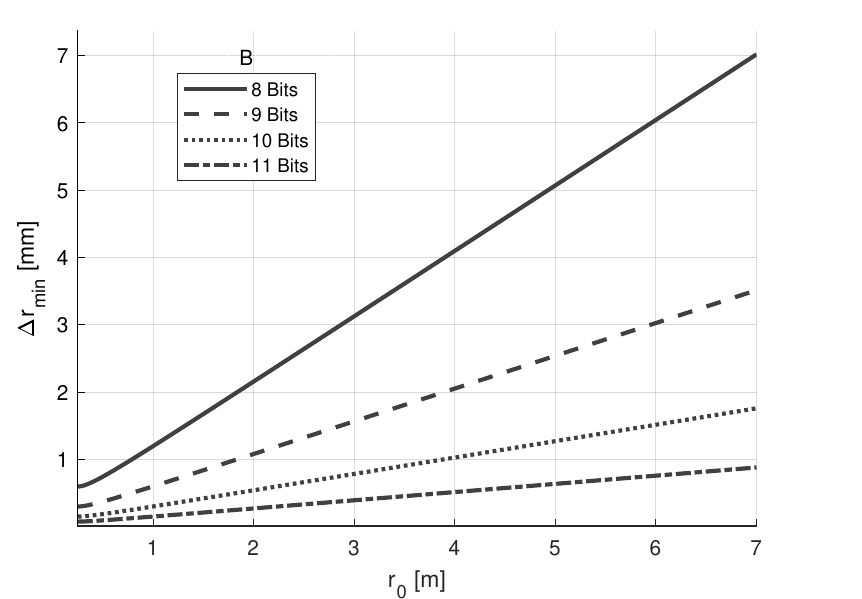}
    \caption{Range resolution $\Delta \rmin$ for common ADC bit values and $\Delta \rmax$ = 50.29 mm.}
    \label{Range_Resolution}
\end{figure}

One may alternatively want to know the best case range resolution for a given LWS system and distance to subject.
Relating equations \ref{Bits_Range_Eqn} and \ref{B_min} to solve for \(\Delta \rmin\) yields:
\begin{equation}
    \Delta \rmin = r_0 - \left[ \frac{(r_0-\Delta \rmax)^{-4}}{2^{\Bmin}} + r_0^{-4} \right] ^ {-\frac{1}{4}}.
\end{equation}
Assuming the same \(\Delta \rmax\) as before, the range resolution is plotted in Figure \ref{Range_Resolution}.
Several important LWS properties can be interpreted from this graph.
First, for ranges beyond half a meter, the capability of any given LWS system to distinguish minimum breathing depth \(\Delta \rmin\) degrades in almost a linear fashion with range. 
However, utilizing a detection scheme with greater number of bits in the ADC can improve the range resolution dramatically, which is particularly important at long range.
As we should expect, each additional ADC bit halves \(\Delta \rmin\) so long as the system is not noise-limited.
The bit requirements for capturing shallow breathing depths do not necessitate extravagant ADCs, suggesting that the more pressing hardware consideration is the noise of the detector.

\subsection{Detector Noise Limitation}
Though we are free to use as many bits as we please, the largest \(B\) that is useful is the one corresponding to the dynamic range of the signal itself, which is:
\begin{equation}
    \Bmax = \log_{2}\left(\frac{\Vmax}{\Vnoise}\right).
\end{equation}

Under this noise limit, the minimum measurable \(\Delta \rmin\) is considered to be the one that produces an output signal with amplitude equal to the noise floor \(\Vnoise\).
This condition is equivalent to a signal-to-noise ratio \(\SNR = 1\) .
While there is noise contributed by the LEDs and lock-in amplifier, the principle source of noise is the photodiode.
The thermal noise variance $\varT$ in the current of a \textit{p-i-n} photodiode is given in \cite{Agrawal} as:
\begin{equation}
    \varT =  \frac{4 \kB T \Fn \Delta f}{\RL}.
\end{equation}
where \(\kB \approx 1.38 \times 10^{-23}\) J/K, \(T\) is the temperature, \(\Fn\) is the detector noise figure, \(\Delta f\) is the detector effective noise bandwidth, and \(\RL\) is the effective load resistance of the transimpedance amplifier.
The shot noise variance is :
\begin{equation}
    \varS =  2 q (\Rd \Pin + \id) \Delta f .
\end{equation}
where \(\id\) is the dark current of the photodiode, \(\Pin\) is the total incident optical power (predominantly ambient light), and the fundamental charge \(q \approx 1.6022 \times 10^{-19}\) C. 
A common approximation for \textit{p-i-n} detectors assumes Gaussian statistics for both the thermal and shot noise contributions \cite{Agrawal}, meaning the total noise variance can be distilled into a single parameter:

\begin{equation}
    \sigma^2 =  \sum_{n} \sigma_n ^ 2.
\end{equation}

The SNR at the ADC input can be expressed in terms of the signal photocurrent \(\Delta \irx\) and the total RMS noise current \(\sigma\) as:
\begin{equation}
    \SNR = \frac{ \Delta \irx }{ \sigma } = \frac{ \Rd \Delta \Prx }
    { \left [ 2q( \Rd \Pin + \id) \Delta f + \frac{4 \kB \Fn \Delta f}{\RL} \right ] ^ {\frac{1}{2}} }.
\end{equation}

Note that the SNR will increase with a smaller \(\Delta f\).
The lock-in amplifier serves to reduce this bandwidth by filtering to accept a narrow frequency window around the modulation rate, thereby improving signal quality.
Setting \(\SNR = 1\), \(\Delta \Prx\) becomes the minimum detectable change in received optical signal power:
\begin{equation}
    \Delta \Pmin = \Prx(r_0-\Delta \rmin) - \Prx(r_0),
\end{equation}
which reduces to
\begin{equation}
    \Delta \Pmin =  {\Rd}^{-1} \left [ 2q( \Rd \Pin + \id) \Delta f + \frac{4 \kB T \Fn \Delta f}{\RL} \right ] ^ {\frac{1}{2}}.
\end{equation}

The range resolution \(\Delta \rmin\) is then determined numerically to be the one which produces the change \(\Delta \Pmin\) for a given LWS configuration.

\section{Experimental Range Resolution}
\label{Range_Resolution_Discussion}
The ADC-limited range resolution is the amount of chest translation required to alter the digital LWS signal by one discrete level.
Owing to the linearity of the LWS signal with $\Delta r$ surrounding a given $r_0$, the range resolution can be calculated from measured LWS data by dividing the physical length of the chest translation by the total number of discrete levels spanning the resulting signal.
Slowly ramping the chest position of the breathing phantom, our LWS system measures a corresponding ramp signal from which both the ADC-limited range resolution and RMS noise can be extracted.

When used to its full potential, our lock-in amplifier outputs digital data with precision corresponding to about 14 bits of dynamic range.
Using 14-bits, it is theoretically possible to represent a displacement of 10 µm with the subject positioned at 0.5 m.
However, in our system, the ADC-limited range resolution at 0.5 m is 36 µm.
This range resolution corresponds to a theoretical 12-bit LWS system.
More precisely, the dynamic range of the measured signal is 1.848 bits less than the fundamental limit predicts. 

\begin{figure}
    \centering
    \includegraphics[clip, trim=0.5cm 0cm 0.3cm 0.5cm, width = \linewidth]{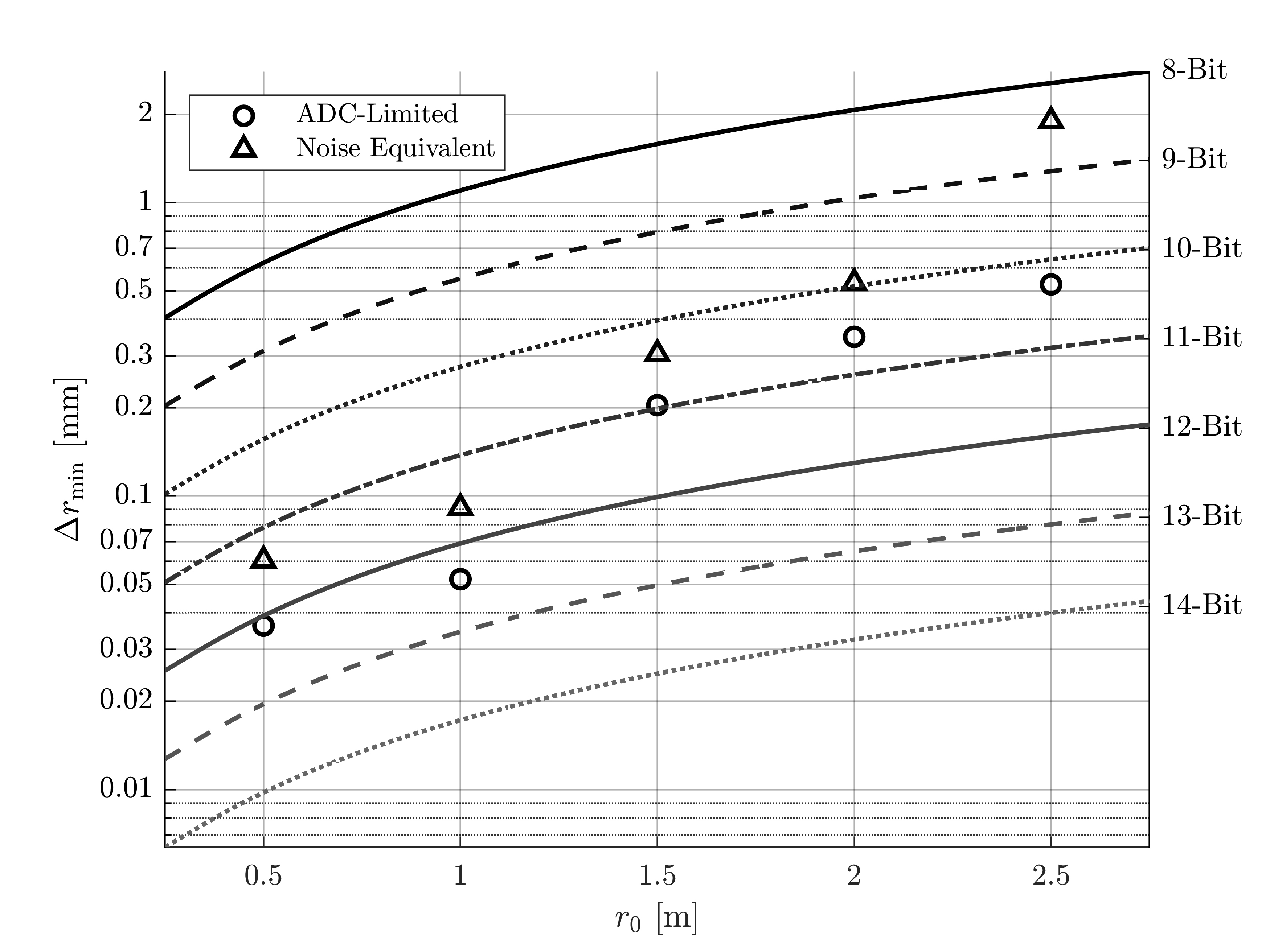}
    \caption{Experimental ADC-limited range resolution and noise equivalent range resolution with expectations from the LWS model. The theoretical traces are labeled by their bit values on the right vertical axis.}
    \label{Noise_Limited}
\end{figure}

Breathing displacements that produce signal variations below the noise floor do not constitute reliable measurements.
Therefore, we define the noise equivalent range resolution as the breathing displacement that produces a signal change equal to the RMS noise voltage.
The noise equivalent range resolution of our LWS system at 0.5 m is 60 µm.
This performance resembles the theoretical resolution for 11-bits of dynamic range. This range resolution is substantially less precise than that of the theoretical limit, suggesting that 3 bits of our lock-in amplifier are not information-carrying at this $r_0$, or that a less capable ADC would be adequate without further limiting the system.
Our experimental ADC-limited and noise equivalent range resolutions are reported in Figure~\ref{Noise_Limited} accompanied by the theoretical resolutions computed for $\Delta \rmax = 30$ mm.
For each measured distance, the noise equivalent range resolution exceeds the ADC-limited resolution.
This indicates that our LWS system is noise limited in typical office lighting conditions at all observed distances.
This reinforces the expectation that noise management is more pivotal to LWS design than the number of ADC bits.

The ambient noise from the photodetector has approximately the same magnitude for any $r_0$, but the LWS signal strength diminishes as $r_{0}^{-4}$.
Accordingly, the noise-limited range resolution of the LWS signal degrades with distance more rapidly than the ADC-limit at long distances.
By 2.5 meters, the noise equivalent $\Delta \rmin = 1.89$ mm, corresponding to a theoretical dynamic range of merely 8.5-bits.
This $\Delta \rmin$ is about the same amount of translation as shallow breathing for some individuals.
Hence, beyond 2.5 meters the respiration rates of certain people are irrecoverable with our system.

At its peak performance, our LWS system has a range resolution $\frac{1}{3}$ as precise as the theoretical limit predicts.
The disparity between the theoretical and the measured ADC-limited resolution widens as $r_0$ is increased.
Some discussion is clearly needed to explain the loss of resolution.
In the theoretical limits, we have assumed that the lock-in amplifier sensitivity can be varied freely so as to equate the maximum chest translation to the maximum output voltage of the amplifier.
The lock-in amplifier used in these experiments has fixed and discrete gain settings that are non-ideal.
As a result, the LWS signal does not occupy the full output range.
For instance, at 0.5~meters, the maximum LWS voltage was 7.65~V, whereas the lock-in amplifier’s maximum output level is 10~V.
This imperfect range utilization represents a loss from theoretical dynamic range of $-0.386$~bits.
Compensating for non-optimal gain selection, there remain 1.462 bits of missing dynamic range for which we must account.

During range resolution measurements, the breathing phantom was situated upright with meters of empty space behind it and to its sides to mimic the hypothetical environment of the fundamental limit calculations.
However, our LWS setup is not perfectly free from clutter.
In addition to the breathing phantom, the light from the modulated source is scattered by the floor, the tripod supporting the breathing phantom, and (to a lesser degree) the walls and other objects in the far extents of the room.
The static component of the LWS signal which results from these unwanted scatterers is known as the clutter signal.

If the breathing phantom is close to the LWS system, the phantom interacts with much of the power from the IR array and occupies a wide patch of the photodetector FOV. 
If the breathing phantom is measured farther from the LWS system, it shadows less of the background.
More modulated light reaches the floor, the ceiling, and other miscellaneous scatterers.
Simultaneously, the clutter occupies proportionately more of the photodetector FOV.
Not only does the relative magnitude of ambient noise increase with distance, but so does the static component of the LWS signal corresponding to clutter.
Clutter reduces the fraction of the output range spanned by the breathing signal.
Hence, the gap between theoretical and measured ADC-limited resolution widens with distance, as evidenced in Figure~\ref{Noise_Limited}.

Scattering from the breathing phantom itself also contributes to clutter, since there are portions of the torso that remain essentially unmoved in the breathing motion.
These conditions create unchanging components of the torso scattered signal that are not encapsulated in the point scatterer model used to establish fundamental limits.
The difference between the breathing phantom’s motion and simple linear translation is an important distinction, which has implications for the LWS range resolution for human subjects as well.
A linear translation of the chest does not necessarily equate to the same translation of the shirt covering.
During range resolution measurements, a shirt was fit snugly on the breathing phantom, but exhibited some slack.
Slack in clothing reduces the effective amount of chest translation, though by no more than a sixth of a bit in our measurements.
The average slack in human clothing is presumed to be greater than this.
The contributions from environmental clutter are small at 0.5 meters, so a bulk of the 1.462 missing dynamic range bits is presumed to be symptomatic of the over-idealistic chest model. 

The continued degradation of the experimental range resolution with distance is attributed to the escalated relevance of clutter when the phantom is far from the LWS.
Since there are nearly two additional bits of deviance from theory at 2.5 meters (beyond the 1.462 bits from 0.5 meters), we estimate $\frac{3}{4}$ of the signal is clutter at this distance.

Unmitigated reflections from the environment significantly affect the LWS signal.
The extremities of the human body are significant contributors to LWS clutter.
The use of highly directive infrared sources reduces the clutter content of the signal by confining the LWS beam primarily to the torso of the subject.
Controlling the spread of the LWS beam is evidently crucial to approaching theoretical range resolutions over long distances.

\section{Application}
\label{Application Section}
\commentout{
In our experimental system, \(\gti\) is limited by ambient light levels, which can lead to DC saturation of the PDA100A integrated transimpedance amplifier.
In a finalized system, that amplifier could be augmented with a DC-blocking filter, amplifying the modulated signal component only.
This opens up the use of the full input range of the ADC reducing \(\Bmin\).
Full advantage should be taken of beamforming optics to reduce offset levels by confining the bulk of \(\Ptx\) to the torso.
The wavelet transform and peak frequency identification can be performed in under a second, though tens of seconds of acquisition latency are unavoidable, given the exceptionally long oscillatory periods of low respiration rates.
}

Widespread deployment of respiration monitoring is contingent on the affordability of the sensing means.
Once sensors are in place, hospitals can begin accumulating vast stores of respiration data paired with corresponding timestamped medical records.
Accessibility of the detection technology is key to the amassing of data.
In the realm of continuous monitoring, non-contact methods are ideal.
Unintrusive operation supports seamless integration into preexisting workflows, such as ceiling mounted detection in an ambulance or over a hospital bed.
Non-contact monitoring can initiate the moment a patient takes their seat in a waiting room---by the time the patient is admitted for examination, their vitals could be already analyzed, extensively.
This technology need not be placed in medical settings alone.
In-home NCRM is an excellent facet for future telemedicine, granting access to the earliest warning signs of disease, improving the efficacy of rapid responses to dangerous respiratory decay.

LWS is a solution to the non-contact respiration monitoring problem which balances concerns of privacy, interference, safety, and cost, making it an ideal candidate for proliferation through the medical and residential infrastructures.
The electronic devices used in our LWS system are all individually well-studied and parameterized.
Fortuitously, even an optoelectronic lock-in amplifier with an integrated phototransistor array for use with LED sources has been engineered \cite{5504046}.
Wavelet filter banks can be implemented with Finite Impulse Response (FIR) filters for real-time, on-chip rate analysis.
The remaining subsystems and components, such as photodiodes and infrared emitters, are mature, having been highly developed by the fiber optics community.
Combining each of these stages, it is entirely possible to produce an \textit{LWS unit on a single chip.}
The massive body of works surrounding the CMOS process has reduced noise levels, power consumption, and manufacturing costs each to impressively minute figures.
Given the modeling framework of this paper and existing hardware architectures, the synthesis of an LWS integrated circuit is reduced primarily to a tuning problem.
Mass-fabricated, ubiquitously embedded into ceilings, walls, desks, steering wheels, and computer screens, the cost-effective potential of LWS can be completely actualized.

\section{Conclusion}
\label{Conclusion Section}
Our findings implicate LWS as a highly viable means of accessing the richness of the health-indicators embedded in respiratory motion.
The nearly indistinguishable errors in the stochastic respiration analysis of subsection \ref{Accuracy Subsection} and accurate respiration rate estimates provide ample proof of the potential of LWS while also revealing its range limitations.
The LWS modeling framework developed in section \ref{LWS_Model_Section} establishes a general engineering foundation and a way to represent LWS systems in simulation.
For instance, one may use the LWS model to query the impacts of different body sizes, attempt to improve detection results by using multiple LWS systems together, or study the effects of clutter and limb motion.
By applying idealized assumptions to the LWS model, this work has also determined the fundamental bounds on range resolution.
Taking practical and analytical limits into consideration, LWS is a cost effective technology for non-contact respiration monitoring in the range of several meters, as our experimental demonstrations have verified. 
Light-Wave Sensing provides a convenient alternative to existing non-contact respiration monitoring technology---it is uniquely fit for wide adoption as a practical aid for the prevention of much suffering.

\bibliography{LWS.bib}
\bibliographystyle{IEEEtran}

\begin{IEEEbiography}[{\includegraphics[width=1in, height=1.25in,clip, keepaspectratio]{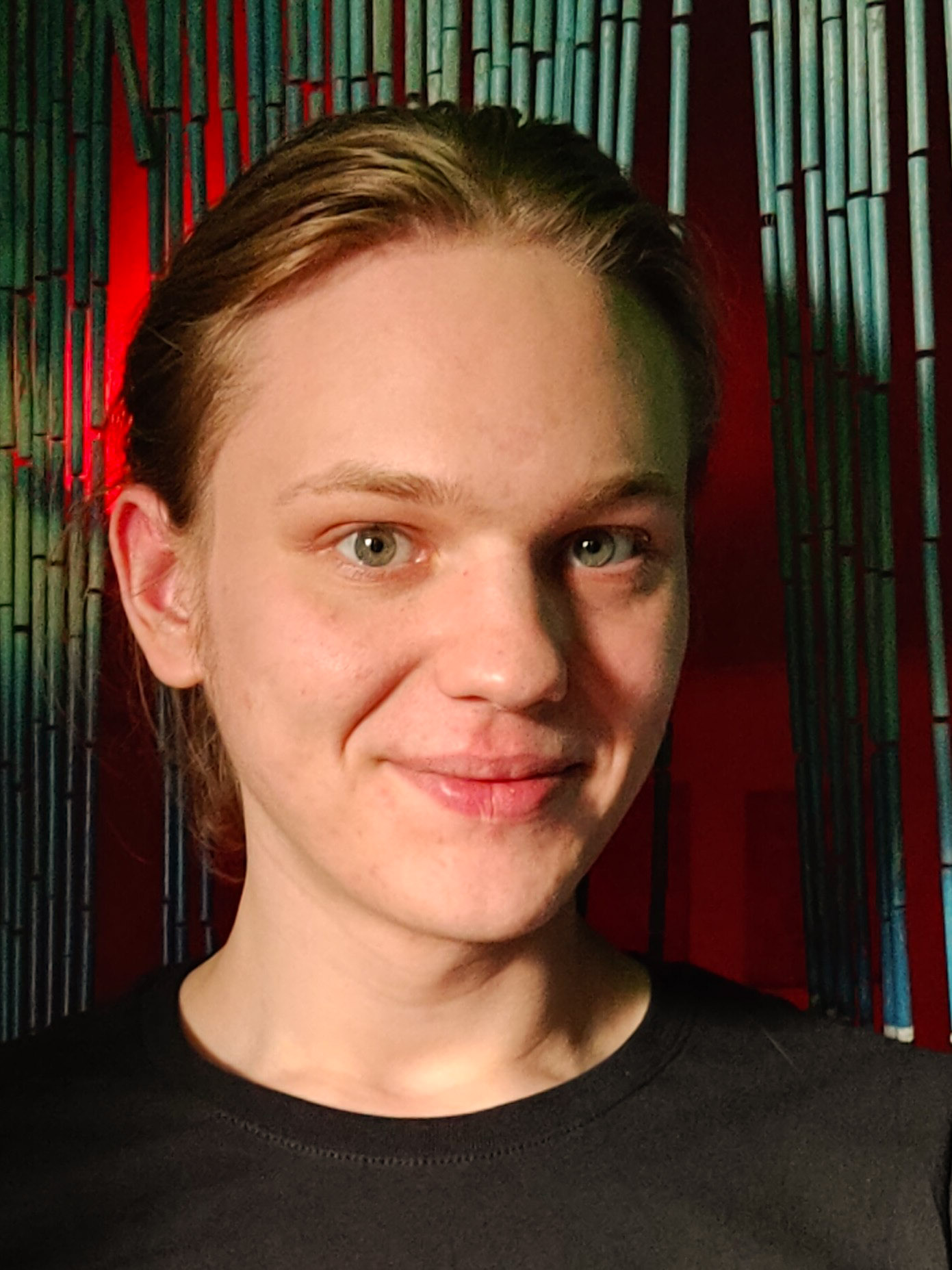}}]{Brenden Martin}
was born in Muskogee, Oklahoma, in the summer of 1999. There he spent his childhood programming, drawing, playing musical instruments, and disassembling electronics. Eventually, Brenden got tired of only taking electronics apart and began learning to put them back together. In 2021, he received his BS in Electrical Engineering from Oklahoma State University (OSU). Brenden now conducts his PhD studies at OSU's Ultrafast Terahertz Optoelectronics Laboratory (UTOL), where he has been involved since he eagerly became an undergraduate researcher as a freshman in 2017. In the UTOL, Brenden’s love of electronics and optics exploded into an ever more obsessive zeal for applied physics. His research interests have come to include ultrafast optoelectronics, optical computation, holography, metamaterials, materials science, and condensed matter physics—although, he is known to show unofficial interest in most any physical phenomenon.
\end{IEEEbiography}

\begin{IEEEbiography}[{\includegraphics[width=1in, height=1.25in,clip, keepaspectratio]{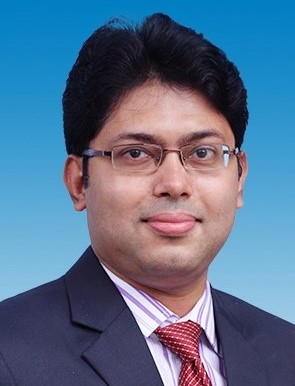}}]{Md Zobaer Islam}
received his B.Sc. degree in Electrical and Electronic Engineering in 2012 from Bangladesh University of Engineering and Technology, Dhaka, Bangladesh. He joined Oklahoma State University, Stillwater, OK as a graduate teaching and research assistant to pursue his Ph.D. degree at the School of Electrical and Computer Engineering in Spring 2020. He has industry experience of 4 years at Bangladesh Telecommunications Company Ltd. In telecommunication and information technology (IT) sector and 3 years at Samsung R\&D Institute Bangladesh in software sector. His current research interests include wireless light-wave sensing and machine learning.
\end{IEEEbiography}

\begin{IEEEbiography}[{\includegraphics[width=1in, height=1.25in,clip, keepaspectratio]{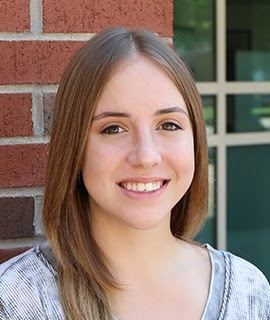}}]{Carly Gotcher}
    is currently working toward her BSEE at the School of Electrical and Computer Engineering, Oklahoma State University (OSU). Carly is presently an undergraduate researcher at OSU’s Robust Electromagnetics Field Testing and Simulation lab (REFTAS), where she pursues her passion for RF-electronics. Previously, Carly has done undergraduate research at the Ultrafast THz Optoelectronic Laboratory (UTOL) and OSU Wireless Laboratory (OWL) including terahertz material characterization and respiration monitoring studies.
\end{IEEEbiography}

\begin{IEEEbiography}[{\includegraphics[width=1in, height=1.25in,clip, keepaspectratio]{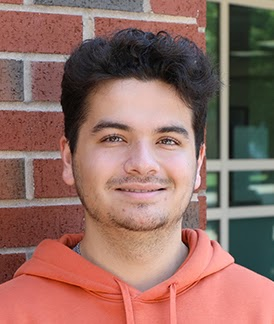}}]{Tyler Martinez}
is an undergraduate student, studying for his BSEE at the School of Electrical and Computer Engineering at Oklahoma State University (OSU). He worked as an undergraduate researcher in the OSU Wireless Lab (OWL) and Ultrafast THz Optoelectronic Laboratory (UTOL) at OSU. Tyler is primarily interested in power systems engineering, especially the diverse methods of energy storage and utilization in development for the future’s smart grid.
\end{IEEEbiography}

\begin{IEEEbiography}[{\includegraphics[width=1in, height=1.25in,clip, keepaspectratio]{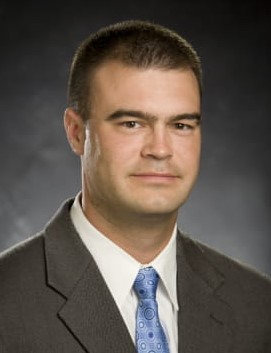}}]{John F. O’Hara}
received his BSEE degree from the University of Michigan in 1998 and his Ph.D. (electrical engineering) from Oklahoma State University in 2003. He was a Director of Central Intelligence Postdoctoral Fellow at Los Alamos National Laboratory (LANL) until 2006. From 2006- 2011, he was with the Center for Integrated Nanotechnologies (LANL) and worked on numerous metamaterial projects involving dynamic control over chirality, resonance frequency, polarization, and modulation of terahertz waves. In 2011, he founded an IoT, automation, and consulting/research company, Wavetech, LLC. In 2017, he joined Oklahoma State University as an Assistant Professor, where he now studies IoT, metamaterials, terahertz communications, and photonic sensing technologies. He has around 100 publications in journals and conference proceedings.
\end{IEEEbiography}

\begin{IEEEbiography}[{\includegraphics[width=1in, height=1.25in,clip, keepaspectratio]{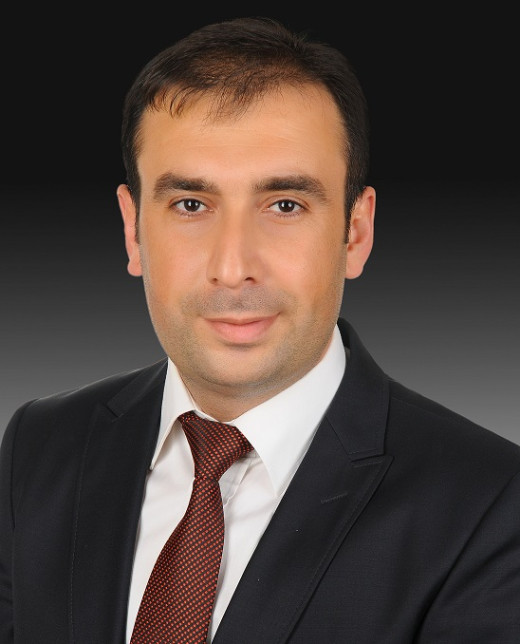}}]{Sabit Ekin}
received his Ph.D. degree in Electrical and Computer Engineering from Texas A\&M University, College Station, TX, USA, in 2012. He has four years of industrial experience as a Senior Modem Systems Engineer at Qualcomm Inc., where he received numerous Qualstar awards for his achievements and contributions to cellular modem receiver design. He is currently an Associate Professor of Engineering Technology and Electrical \& Computer Engineering at Texas A\&M University. Prior to this, he was an Associate Professor of Electrical and Computer Engineering at Oklahoma State University. His research interests include the design and analysis of wireless systems, encompassing mmWave and terahertz communications from both theoretical and practical perspectives, visible light sensing, communications and applications, noncontact health monitoring, and Internet of Things applications.
\end{IEEEbiography}

\end{document}